\documentclass[12pt]{article}
\usepackage{color,graphicx}
\usepackage[utf8]{inputenc}

\newcommand{\beq}{\begin{equation}}
\newcommand{\eeq}{\end{equation}}
\newcommand{\beqs}{\begin{eqnarray}}
\newcommand{\eeqs}{\end{eqnarray}}

\def \be {\begin{equation}}
\def \ee {\end{equation}}
\def \bea {\begin{eqnarray}}
\def \eea {\end{eqnarray}}
\def \nn {\nonumber}

\def \lab #1 {\label{#1}}

\definecolor{agr}{rgb}{0.0,0.4,0.0}
\definecolor{apu}{rgb}{0.3,0.2,0.6}
\definecolor{dre}{rgb}{0.6,0.0,0.1}

\ifx\pdfoutput\undefined
\input epsf

\def\figscale#1#2{\epsfxsize=#2\epsfbox{#1.eps}}
\else

\def\figscale#1#2{\pdfximage width#2 {#1.pdf}\pdfrefximage\pdflastximage}
\fi

\def \bea{\begin{eqnarray}}
\def \eea{\end{eqnarray}}
\def \ba{\begin{eqnarray}}
\def \ea{\end{eqnarray}}

\def\figscale#1#2{\pdfximage width#2 {#1.pdf}\pdfrefximage\pdflastximage}

\def \nn{\nonumber}

\begin{document}

\begin{center}
\Large
{\bf Comments on Collinear Factorization}\footnote{contribution to Snowmass 2021}
\end{center}

\medskip

\begin{center}
{ {\bf George Sterman}\\[5mm]
C.N. Yang Institute for Theoretical Physics and Department of Physics and Astronomy\\
Stony Brook University, Stony Brook NY 11794-3840 USA\\
george.sterman@stonybrook.edu
}
\end{center}

\medskip

\begin{abstract}
In this short review, I'll discuss  the background, applicability and prospects of collinear factorization in quantum chromodynamics.
\end{abstract}


\tableofcontents


\section{Introduction: some concepts of factorization}

Factorization is the fulcrum 
of the collider system, which, realizing the famous quote of Archimedes, is designed to lift new worlds from the vacuum, with 
high-energy colliding protons providing the lever arm.  It is central to the energy frontier program, entering into nearly every attempt to achieve precision within
the Standard Model, and in recognizing signals of new physics \cite{Heinrich:2020ybq,Caola:2022ayt}.   As our searches become more sophisticated,
they become more sensitive to the dynamics of the final state.   Such sensitivity, however, may violate the
assumptions necessary to prove factorization.   The purpose of this little essay is to present one person's viewpoint on 
the basic concepts that have gone into proofs of  collinear factorization at hadron colliders,  the
fundamental results, and some of their strengths and limitations, 
to suggest where we can be confident, and where more work remains to be done.  The presentation is
a mixture of intuitive and technical arguments, hopefully in a conversational style.  It is thus by no means intended
to provide a final word, or to be rigorous, but rather to lead to further conversation as the field moves forward into the next era of
collider physics.

 Factorization is the quantum field theoretic realization of the parton model \cite{Feynman:1969ej,Bjorken:1969ja}, in which large momentum transfer cross sections in hadron-hadron (or lepton-hadron) collisions are the result of  the short-distance collisions of constituents within those  hadrons, each Lorentz-contracted and time dilated in the frame of the other.  The word ``factorization" in this context refers to our ability to treat the momentum distributions of the partons as incoherent from the probabilities of their short-distance interactions, if and only to the extent that they collide at relative velocities approaching the speed of light.  In this limit, before the collision each hadron is in a superposition of quantum states that is completely independent of the other hadron.   At high enough energies, all hadrons are ``pristine" at the moment of collision, in the same mixture of states since baryogenesis.  The basic concept here is {\it causality}, in the sense that signals from one hadron cannot reach the other before the moment of collision.   This implies universality of the kinematic distributions of constituents in fractional momenta, the parton distributions.  In collinear factorization, the hard scattering probability function is computed with the momenta of the colliding partons taken lightlike and in the directions of the colliding hadrons.   
  
 The collision takes place over a short time scale, creating new particles and/or patterns of momentum flow that evolve into the observed final state.   Interactions that take place  long after the short-distance process involve smaller momentum transfers, which for suitably defined observables do not alter the overall flow of momentum, or excite quantum numbers in a manner that can neutralize those created the hard scattering.   When summed with sufficient inclusivity, perturbative singularities associated with this evolution cancel, because they do not affect the total probability that defines the set of states that are summed.   This is an expression of the {\it unitarity} of the underlying theory.   Here,  the identifiable class of final states is typically characterized by decay products of transient, heavy degrees of freedom and/or high-$p_T$ leptons or jets.     These may be Standard Model or not, and final states may be visible or dark, so long as we can recognize their presence, directly or indirectly.

In summary, the central  concepts of factorization are causality, unitarity and universality.    Most commonly, universality refers to parton distributions, but in a closely related sense universality also applies to hard scattering functions. For parton distributions, it applies between different processes, say the proton distributions in deep-inelastic scattering for an electron-proton initial state, and in Drell-Yan, or jet production production for hadron-hadron scattering.    For the short-distance hard-scattering factors, we assume a universality among perturbative calculations with different long-distance completions of the theory.  For example, the production of a photon plus gluon from a quark pair is the same in real QCD as in the four-dimensional limit of dimensionally-continued QCD.    The universality of parton distributions allows us to transfer information from deep-inelastic scattering to hadron-hadron scattering and vice-versa.   The universality of short-distance factors allows us to compute hard scattering for incoming partons rather than model hadrons, clearly a much more efficient procedure.

In principle, as we go to higher orders in hard scatterings, simultaneously fitting more precisely determined parton distributions within processes we confidently identify as Standard Model at short distances, we can achieve more precision in the comparison of high-statistics data to our calculations, with an eye out for discrepancies.   At the same time, factorization allows us to calculate with the same precision the ways that new forces and fields would manifest themselves in these same data.  In the best of possible worlds, we may see behavior that is qualitatively different from the Standard Model predictions (asymmetries, etc.)   In all this, uncertainties in data, and in parton distributions are subjects of increasing interest.
   
Collinear factorization is well-illustrated by
the inclusive cross section for the production of a Drell-Yan pair or more generally, a heavy electroweak boson of 
mass $Q$,
\bea
Q^4 \frac{d\sigma^{pp\to \gamma^*(Q)X}}{dQ^2 }\,=\, \sum_{ab}\int dx_a \,f_{a/p}(x_a) \int dx_b \,f_{b/p}(x_b) 
\,\omega^{ab\to \gamma^*(Q)X}(x_aP_A,x_bP_B,Q)\,,
\label{eq:dy-fac}
\eea
with the functions $f_{i/p}(x_i)$ the parton distributions, and $\omega^{ab\to \gamma^*(Q)X}(x_aP_A,x_bP_B,Q)$ the perturbatively
calculable short-distance functions for partons $a$ and $b$ to produce an off-shell photon of mass $Q$.
Similarly, jet cross sections take the generic collinear-factorized form
\bea
S\, \sigma^{pp\to \{j_i\}X}\,=\, \sum_{ab}\int dx_a \,f_a(x_a) \int dx_b \,f_b(x_b) 
\,\omega^{ab\to \{c_i(p_i)\}X} (x_aP_A,x_bP_B,\{p_i,\delta_i\})\, .
\label{eq:jet-fac}
\eea
where $\{p_i,\delta_i\}$ labels the momenta of the observed jets and the paramters $\delta_i$ that define them, jet cone radii, for example.  (Notice that these parameters generally imply limitations on the additional particles labelled by $X$.)
In both cases, the short-distance, perturbative function, $\omega$, is normalized to be dimensionless.   We have suppressed
the factorization scale, which for this discussion we assume is of the order of the scale of renormalization.
Clear generalizations exist to multiple electroweak boson production followed by decay, and to an unlimited set of
``new physics" scenarios, as long as these involve heavy particles and short-distance interactions.

\section{What's involved in factorization proofs}

As suggested above, the basic ideas that go into arguments for factorization are causality and unitarity, with universality as a consequence.   In informal language, the formation of initial states takes place too early to affect the hard scattering, and subsequent interactions take place to late to undo it.   Also as hinted above, the applicability of these ideas very much depends on what we measure in the final states, and how we combine them.
 
 First a word about what is meant by {\it proof} in this context.   All proofs I know of assume, as noted above, that hard-scattering functions are universal, and that they can be identified in the scattering of elementary quanta, generally in a theory that is infrared-regulated dimensionally or by other means.   In that theory, long-distance dynamics is to be identified through its (regulated) infrared singularities, generically poles in $4-D$, with $D$ the number of dimensions.  When a cross section is factorized in this theory, we can identify the short distance function by a suitable definition and separation of parton distributions, which contain all infrared singularities.   In this procedure, our ability of factorize depends on our ability to identify singular long-distance, infrared (IR) behavior in the massless limit.   Once identified, we investigate whether in each relevant region of phase and loop momentum space, the integrand is either consistent with factorization, or is part of a nonfactorizing class of contributions that cancels without double counting.  In all these arguments, a key tool is that physical infrared singularities of all types are logarithmic in gauge theory cross sections.  As a result, a single power suppression in a sum over logarithmically-divergent integrals leads to an integral that is finite in perturbation theory.   Effective theory treatments organize sets of regions where logarithmic singularities arise from the outset.  The consistency of any such choice is verified in the process of matching to the full theory at each order of perturbation theory at relevant scales.   The success of this matching at all orders requires in general arguments that go beyond the effective theory itself, and are part of its justification.
 
 For renormalizable, non-gauge theories near four dimensions, like $\phi^4$ and Yukawa, or $\phi^3$ near six dimensions, IR  singularities are all ``collinear", corresponding to splittings and recombinations of descendants of the two incoming partons, and the task just described is by now well understood.   The treatment of collinear singularities in gauge theories, it is worth noting, also requires the treatment of unphysical, longitudinally-polarized vector particles that share the momenta of active partons, at least in covariant gauges.   Many of the early all-orders discussions of factorization avoided this complication by working in axial or other physical gauges.   A dedicated discussion of the role (and factorization) of unphysical collinear gluons is given, for example, in Ref.\ \cite{Collins:1989gx}.    In effective theory treatments, unphysical gluon polarizations can be absorbed into the definitions of the physical fields \cite{Stewart:2009yx,Becher:2014oda}. 
 
 For gauge theories, an additional distinguishing feature is soft singularities.  In essense, soft gluons can be exchanged between constituents of the incoming hadrons, and such exchanges appear to have the potential to violate causality, and hence factorization.\footnote{Actually, the violation of factorization would be the least of our problems.}    Roughly speaking, it's as if the quarks in proton A could attract the antiquarks in proton B before they collide, even if they approach each other at the speed of light.  They do not do so, because until the collision the fields involved are gauge-equivalent to zero field \cite{Basu:1984ba}, but this feature is obscured in  perturbation theory, which incorporates sums over physical and unphysical degrees of freedom.    Factorization proofs rely on our ability to organize the phases associated with the external charged particles.   The challenge is to disentangle the unphysical component of these phases from the physical gluon-mediated scattering, and to discover when and how the phases cancel in cross sections.   At this stage, we emphasize again that all these arguments depend crucially on treating the momenta of incoming particles, the partons and hadrons, as lightlike.   
 
Early all-orders discussions of factorization in hadron-hadron scattering \cite{Sachrajda:1978ja,Amati:1978wx,Amati:1978by,Ellis:1978ty,Libby:1978qf,Efremov:1978xm} dealt most exhaustively with collinear singularities, which control evolution, also providing a variety of arguments for the cancellation of soft singularities through virtual-real emission cancellation.
This work was inspired in part by a pioneering low-order demonstration of  soft-gluon cancellation in Ref.\ \cite{Politzer:1977fi} and the older $\phi^4$ fragmentation analysis in Ref.\ \cite{Mueller:1974yp}.  In what follows, I will give an informal account of the soft gluon interactions in hadronic hard scattering, which set QCD and other gauge theories apart, and were the primary concern of a second round of factorization studies
\cite{Collins:1983ju,Bodwin:1984hc,Collins:1985ue,Collins:1988ig}.\footnote{For a relatively brief review of the basic arguments, see \cite{Collins:1998ps}, and for a much more complete discussion, Ref.\ \cite{Collins:2011zzd}.}

 \subsection{Soft gluons and phases}

Associated with each parton that arrives at the hard scattering from the past is a nonabelian phase, sensitive only to that component of the gauge field that contracts with the momentum of the parton.    
Such interactions can be organized into a phase by the application of an appropriate gauge transformation. Likewise, for the particles that emerge from the hard scattering, their experience of the color fields of other emerging particles is entirely pure gauge, again involving only one component of the gauge field.   These phases differ in an important way.    This difference illustrates the fundamental relationship between causality and analyticity in quantum field theory, and is a direct consequence of the kinematics of the process.
The interactions of fast-moving partons over a semi-infinite  time scale, $\lambda$, take one of the two general forms, in terms of a ``current", labelled here as simply $J$,
\bea
\tilde J^\mu_+(k\cdot \beta) \ &=&\ \beta^\mu \int_0^\infty d\lambda\ e^{ i \lambda k\cdot \beta} J(\lambda\beta)\, ,
\nn\\[2mm]
\tilde J^\mu_-(k\cdot \beta) \ &=&\ \beta^\mu \int_{-\infty}^0 d\lambda\ e^{ i \lambda k\cdot \beta} J(\lambda\beta)\, ,
\eea
where $\beta^\mu=p^\mu/p^0$ defines the light-like velocity of the parton.
The vector $\tilde J^\mu_+(k\cdot\beta)$ is analytic in the upper half-plane in $k\cdot \beta$, simply because, for Im($k\cdot\beta$) and $\lambda$ both positive, 
the integral that defines $\tilde J^\mu_+$ becomes exponentially suppressed for large $\lambda$.  Similarly,
$\tilde J^\mu_-(k\cdot\beta)$ is analytic in the lower half-plane of $k\cdot\beta$.

An essential feature of factorization proofs is the wholesale continuation of momentum space contour integrations away
from singularities, into the regions of analyticity just identified.  Along the deformed contours, the ``eikonal approximation", $p\cdot k\gg k^2$, holds for 
generic soft momentum $k^\mu$ and high-energy parton momentum $p^\mu$.   This condition, in turn, implies that
in a subdiagram consisting of collinear lines whose momenta are all of order $p$, all components of momentum
$k^\mu$ can be neglected in the analysis of singularities, {\it except} for $\beta\cdot k$.  This is because a single factor like $k^2/(\beta\cdot k)$ renders
a logarithmically divergent integral finite. For lines that terminate at the hard scattering
(``active" incoming partons) or originate at the hard scattering (produced high-$p_T$ partons), we can argue that no additional singularities prevent this deformation \cite{Collins:1989gx} and that all potentially ``acausal" effects
are organized into unitary Wilson line operators (nonabelian phase operators), coming in from the past to the origin, with lightlike 
velocities $\beta_i$, or going out to the future with lightlike velocities $\beta_j$,
\bea
\Phi_{\beta_i,{\rm in}}^{(R_i)}{}(x)
\ &=& \
{\rm P}\ \exp \left[ -ig\int_{-\infty}^0\ d\lambda\, 
\beta_i\cdot
A^{(R_i)}(\lambda\beta_i+x)\, \right]\, ,
\nn\\[2mm]
\Phi_{\beta_j,{\rm out}}^{(R_j)}{}(x)
\ &=& \
{\rm P}\ \exp \left[ -ig\int^\infty_0\ d\lambda\, 
\beta_j\cdot
A^{(R_j)}(\lambda\beta_j+x)\, \right]\, .
\eea
Here $R$ is the color representation of the parton in question,  and P denotes path ordering.    For elastic S-matrix elements in wide-angle scattering, these operators
fully describe the interactions of soft gluons, and generate all associated phases \cite{Sen:1982bt,Feige:2014wja,Ma:2019hjq}.  For hard-scattering cross sections, however, the situation is more complicated.

\begin{figure}[t]
\centering
\vspace{10mm}
\centerline{\figscale{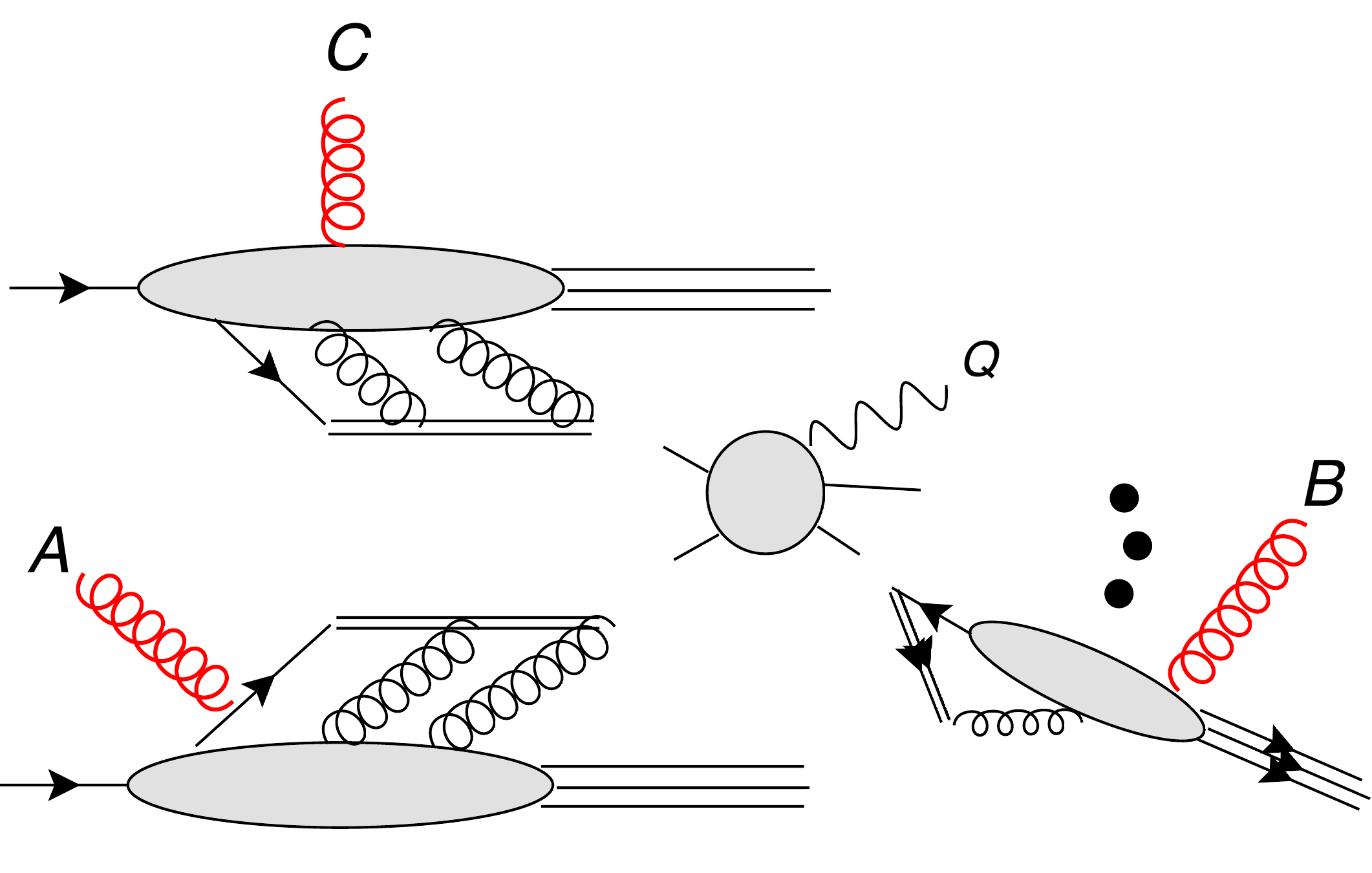}{8cm}}
\caption{Representation of components of a hard-scattering amplitude in terms of two incoming and a representative outgoing jet in a typical hard-scattering process.  The labelled gluons attach to (A) an incoming active parton, (B) a parton emerging from the hard scattering and (C) a spectator parton.  Double lines represent the couplings of the unphysically-polarized collinear gluons referred to in the text.  \label{fig:amplevel}}
\end{figure}

 Referring to Fig.\ \ref{fig:amplevel}, 
 integrals over soft gluon attachments to spectator lines can extend from negative to positive infinity on the same line, and hence inolve both $\Phi_{\rm in}$ and $\Phi_{\rm out}$ operators.  
 This leads to a mismatch in analyticity properties between gluons in the initial state and those in the final state, and this is a physical effect.
 This distinction cannot be removed by a gauge transformation, and corresponds physically to gluons that arrive at the time of the hard scattering.  
 Such gluons are the origin of the ``Glauber", or ``Coulomb", gluon problem in hadron-hadron scattering \cite{Collins:1981ta,Bodwin:1981fv}.   In essence, Glauber gluons occur in regions of momentum
 space where integration contours are pinched between singularities associated with incoming and outgoing phase operators, hence for which the eikonal approximation fails.
If such a gluon also attaches to the oppositely moving incoming line, we have an essential obstacle to factorization.   Corrections of this type only begin at three loops at the finite level,
with logarithmic enhancemennts starting at four loops.

It is at this stage that the construction of the cross section comes in.   For appropriately defined cross sections, final-state phase operators (the $\Phi_{\rm out}$s)
cancel in the sum over final states.   

The purest example of an outgoing phase is in  deep-inelastic scattering, where it is generated by soft gluon interactions with the outgoing quark (and its collinear radiation).   This is the origin of the matrix element definition of parton distributions, which can be written for quark field $q_a(x)$ with flavor index $a$, in the conventional form \cite{Collins:1981uw}
\bea
f_{a/h}(x,\mu)\ =\ \frac{1}{4\pi}\, \int_{-\infty}^\infty d\lambda e^{-i\lambda n\cdot (xp)}\ \left \langle h(p) \left | 
\bar q_a (\lambda n)\, \Phi^\dagger{}_{n,\rm out}(\lambda n)\, \frac{n\cdot \gamma}{2}  \Phi_{n, {\rm out}}(0)   q_a(0) \right |h(p) \right \rangle\, .
\label{eq:pdf}
\nn\\
 \eea
 Here, soft singularities cancel in the final state because of the unitarity of the Wilson line operators, which are both oriented in direction $n^\mu$, conventionally chosen
 as a lightlike direction opposite to the lightlike momentum $p$ of incoming hadron $h$.   Argument $\mu$ is the factorization scale, the scale associated with renormalization of the composite operator that defines the matrix element \cite{Collins:1981uw}, which becomes ultraviolet divergent
 because its fields are all separated by lightlike distances ($n^2=0$).   In effect, the scale $\mu$ cuts off the transverse momenta of radiation.   The transverse momentum can
 actually be fixed in this matrix element.   This possibility is realized in transverse momentum distributions and factorization, with many applications in deep-inelastic scattering and Drell-Yan proceses \cite{Rogers:2015sqa}.

 We can extend collinear factorization in deeply inelastic scattering far beyond the total cross section, to sums designed to identify jet and related final states, because collinear
 and soft singularities cancel in the corresponding partial sums over states.   This is a consequence of the  ``local" unitarity \cite{Sterman:1993hfp,Sterman:1995fz,Capatti:2022tit}  of quantum field theories.
 Unitarity at fixed values of spatial momenta is the basis of the infrared safety of jet cross sections in $\rm e^+e^-$ annihilation \cite{Sterman:1978bi,Sterman:1978bj,Sterman:1979uw},  and is closely related to the ideas of effective field theory, and to efforts to reformulate calculations
 entirely in four dimensions \cite{TorresBobadilla:2020ekr}.   For example, an interaction Hamiltonian that excites and absorbs partons in only two directions is not rotationally-invariant, yet it is hermitian, and hence generates a unitary theory.   As such, summing over its final states gives, by the optical theorem, a forward scattering amplitude that is free of
 infrared  singularities, either soft or collinear.

 \subsection{Unitarity, final states and soft gluons}

Moving on to hadron-hadron scattering, the essential observation is that in inclusive cross sections, final state interactions, including phases cancel in the sum over final states.   This will be the case not only
for particles emerging from the hard scattering at large $p_T$, but also for spectator particles that are collinear to the incoming directions.   It is this cancellation that eliminates pinches in the Glauber regions.

The ``local" unitarity of QCD, mentioned above, enables the cancellation for sectors in momentum space that describe jet cross sections.   
The long-distance phases  that remain are generated only by initial state operators, $\Phi_{\rm in}$.   Given what we've seen, this is somewhat surprising at first, because it is just the opposite to deep-inelastic scattering, built, as in Eq.\ (\ref{eq:pdf}),  on $\Phi_{\rm out}$ operators. Nevertheless,  universal factorization into parton distribution functions, as naturally defined in deep-inelastic scattering, is still possible so long as soft singularities cancel in the hadron-hadron cross section in question.   
 This cancellation was referred to as the transition from ``weak" to ``strong" factorization in the context of Drell-Yan cross sections,  \cite{Collins:1983ju,Bodwin:1984hc,Collins:1985ue,Collins:1988ig}.  Reference \cite{Aybat:2008ct}  discussed explicitly how soft radiation sensitive to the difference between the incoming and outgoing Wilson lines cancels.   
 
 The technical consequence of this cancellation is that the interactions of all ``acausal" initial-state gluons connected to the
set of partons collinear to incoming lines A and B become coherent, and are summarized by two Wilson lines of the type $\Phi_{\rm in}$.   
In a fully inclusive cross section, we expect all soft singularities associated with these operators to cancel, enabling a universal factorization.   This is the case of Drell-Yan processes.   It is natural to ask, however,
about less inclusive cross sections.   How inclusive is inclusive enough, and what sums over final states will cancel outgoing nonabelian phases up to a momentum scale at which perturbation theory can be applied?

Here is one way of thinking about these questions.  We can characterize the length scales associated with the hard scattering by the energy scales of heavy or high-$p_T$ particles in the final state.  This alone, however, does not ensure that final state phases cancel.   Rather, a  ``figure of merit" corresponding to this cancellation is  the  coordinate space distance between the hard scattering in the amplitude and in the complex conjugate.   At zero separation, outgoing Wilson lines cancel identically.   

The relative ``displacement" of the hard scattering is determined by the range of momenta integrated in the final state, by the usual complementarity of position and momentum measurements.   For hard scale $Q$, if we integrate over a range of momenta in the final state characterized by scales $\delta Q$, with $\delta$ a fraction such that $\alpha_s(Q) \ln (1/\delta) < 1$, we anticipate that the hard scatterings are localized on the scale $1/(\delta Q)$.   Such scales can be used to define jet and related cross sections.   In these cases, the final-state phases $\Phi_{\rm out}$ corresponding to outgoing partons will be matched with their hermitian conjugates at scales that are moderately close to $1/(\delta Q)$, and we can expect their soft gluons to cancel up to the scale $\delta Q$, which is consistent with the definition of the cross section.   As $\delta$ decreases, of course, the cross section becomes more and more sensitive to soft radiation.   At whatever scale the non cancellation sets in, it is no longer possible to absorb radiation at that energy scale (or higher) into cancelling nonabelian initial state phases, and radiation is no longer incoherent between the descendants of the incoming particles.   At scales beyond this one, factorization is ``broken", in the sense found for amplitudes with forward collinear lines in the final state \cite{Catani:2011st}.    Perhaps more accurately, what we have is collinear factorization at an anomalously low factorization scale, determined by the parameters of the set of final states, rather than the hard scattering itself.   Beyond this scale, in fact, virtual or real radiation coupling to fragments of the colliding hadrons, the spectators, can influence measured quantities.  As noted above, such corrections, which are consistent with collinear factorization for inclusive cross sections, begin at three loops.    Important examples of cross sections that illustrate the limitations of collinear (and transverse) momentum factorization in cases like these have been identified, for example, in Refs.\ \cite{Forshaw:2012bi,Rogers:2010dm,Zeng:2015iba,Rothstein:2016bsq,Schwartz:2018obd}.   

Given these general considerations, I'll go on in the next section to give a viewpoint on some of the commonly-discussed cross sections, and the arguments I'm aware of that determine their factorization status.

\section{Where we stand}

At the risk of introducing evanescent terminology, I'd like to distinguish between ``deep" factorization, at the scale of the hard scattering, and ``shallow" factorization, in which the factorization scale is set by relatively small (if still perturbative) energy scales that are built into the final state sum in question.    It is important to remember, though, that when such scales decrease, corresponding to a less inclusive cross section, with fewer final states, the cross section must decrease.   To see this decrease, it will be necessary to generalize the analysis of Sudakov resummation, which can be carried out for fixed-angle exclusive amplitudes  \cite{Sen:1982bt,Feige:2014wja,Ma:2019hjq}.   In time, we will learn to love cross sections with shallow factorization, which are in some sense intermediate between inclusive cross sections and exclusive amplitudes.   They encode the coherence of color flow in the perturbative regime, in manners that we have not yet fully understood.

\subsection{Processes with deep factorization}

The arguments above provide, I hope, sufficient framework to interpret at least the claims in the literature of what factorization arguments purport to show.   In these terms, a list of processes for which factorization scales are similar to the corresponding scale of the hard scattering includes electroweak annihilation (Drell-Yan, electroweak boson and Higgs production), and jet production in a wide range of kinematics.   

\begin{itemize}

\item  {\it Electroweak annihilation (fully inclusive in hadronic final states). }  These cross sections, were discussed explicitly in the ``second round" of factorization papers, from the mid 1980's \cite{Bodwin:1984hc,Collins:1985ue,Collins:1988ig}.   In this case, the cancellation of final states is complete, a claim that is made explicit by the use of time-ordered or light cone-ordered perturbation theory.   Later work \cite{Aybat:2008ct} explored more in detail the manner in which soft radiation associated with the initial state phases cancels, leading to classical ``universality" for the parton distributions.   Similar results can hold in multiple hard scatterings, where, for example, two electroweak bosons are created at separate partonic interactions in a single collision \cite{Diehl:2015bca}.

\item  {\it Inclusive jet production}.  Here, we imagine sets of states with jets defined through parameters $\epsilon$ with $\alpha_s(Q) \ln (1/\epsilon) \sim 1$.   For a review, see \cite{Sapeta:2015gee}.   The discussion of jet cross sections began early \cite{Sachrajda:1978ja}, and in Ref.\ \cite{Libby:1978qf} the local unitarity mentioned above was used to show that collinear factorization for jet cross sections of this sort reduces to the Drell-Yan case:  soft quanta that connect active or spectator partons to lines in the outgoing jets cancel in the sum over jet final states (assuming that soft radiation is summed inclusively up to a finite fraction of the jet energy scales).   This leaves only soft radiation connecting the two incoming jets, just as in Drell-Yan.

\item  {\it Single-particle inclusive cross sections}, with inclusivity defined similarly to the jet cross sections defined above, provides another example.   Arguments for the use of universal fragmentation functions  were given in Ref.\ \cite{Nayak:2005rt}, very much relying on the concepts described above.

\item   {\it Multi-particle inclusive cross sections for massive quarks}:   As an example, consider final states with a pair of top quarks,.  We may observe the top momentum with arbitrary accuracy, so long as we average over the momenta of the anti-top \cite{Mitov:2012gt}.
For cross sections with the hard scale far above the mass of a heavy quark, additional considerations enter  \cite{Beneke:2016jpx,Caola:2020xup}.
Interestingly, the same cancellations will occur if we measure the transverse momentum of both elements of the top pair, if the pair recoils against sufficiently hard gluonic radiation, over which we average.   This is possible, of course, only at an order high enough that such radiation is present in the final state.     Thus, we expect the cancellation to fail at low orders, and it does \cite{Mitov:2012gt,Collins:2007nk,Collins:2007jp}.   Nevertheless, in the production of very heavy pairs, the cross section is actually dominated by final states with substantial radiation.   In this case, the nonradiative portion of the cross section is Sudakov suppressed.   In this limit, then, it might be possible to compute the pair transverse momentum self-consistently.

\end{itemize}

\subsection{Shallow factorization}

The generic examples of shallow factorization are nonglobal sums over final states \cite{Dasgupta:2001sh}.   In these cross sections, we impose an energy cutoff in one angular region at a scale much lower than that of unobserved ``jets and subjets" in another region.   In general, at high orders these lead to ``super-leading" logarithmic enhancements \cite{Forshaw:2012bi,Becher:2021zkk}, sensitive to both soft and collinear limits, and the factorization scale decreases to the measure of energy flow, rather than the underlying hard scale.   Again, in this case, we can think of these logs as due to the non cancellation of outgoing nonabelian phases for energy scales larger than the cutoff imposed to define the cross section.  

Going to the limit, if we ``empty out" a section of phase space (say, a complete rapidity gap), we break the cancellation of final-state interactions, including the matching of nonabelian phases in the amplitude and complex conjugate in hadron-hadron scattering.  Nevertheless, in deep inelastic scattering a variant factorization is possible \cite{Collins:1997sr}, which goes beyond the present discussion.


\section{Conclusions}

In this short review, I've tried to summarize some of the elements of the existing literature on collinear factorization, concentrating on basic intuition and the role of soft radiation.
I have left to the side many topics, including  transverse momentum factorization \cite{Rogers:2015sqa}, factorization for generalized parton distributions \cite{Belitsky:2005qn},  the uses of factorization in resummation, higher twist factorization, and its application in nuclear targets and collisions.
Even purely collinear factorization is a vast subject, and in closing I can only emphasize that even within this narrow focus on a familiar but consequential topic, much important work has surely been neglected.
Renewed interest in this subject, partly inspired by the high energy behavior of the spontaeneously broken sectors of the Standard Model \cite{Bauer:2017isx,Han:2020uid,Chien:2018ohd,Baumgart:2018ntv}, will hopefully lead
to further insights in how to use examples of ``shallow" factorization, to study the transition between short- and long-distance dynamics in gauge theories.   

\subsection*{Acknoledgements}  I am grateful to Werner Vogelsang for helpful comments, and to Fabio Maltoni, Shufang Shu and Jesse Thaler for their encouragment of the idea of preparing a short discussion of the status of collinear factorization in connection with the Snowmass process.   This work was supported in part by the National Science Foundation, award PHY 1915093.

\end{document}